\documentclass[a4paper,fleqn,usenatbib]{mnras}
\usepackage{times}

\usepackage{mathtools}
\usepackage{aas_macros}

\usepackage{graphicx}
\usepackage{amstext}
\usepackage{enumitem}

\usepackage{amssymb}

\makeatletter
\newcommand{\mathleft}{\@fleqntrue\@mathmargin0pt}
\newcommand{\mathcenter}{\@fleqnfalse}
\makeatother

\title[Suppressing CBP with a new constraint on overshoot]{The treatment of mixing in core helium burning models -- III. Suppressing core breathing pulses with a new constraint on overshoot}

\author[Constantino et al.]{Thomas Constantino$^{1,2}$\thanks{E-mail: T.Constantino@exeter.ac.uk}, Simon W. Campbell$^{3,2}$ and John C. Lattanzio$^{2}$\\
$^{1}$Physics and Astronomy, University of Exeter, Exeter, EX4 4QL, United Kingdom\\
$^{2}$Monash Centre for Astrophysics (MoCA), School of Physics and Astronomy, Monash University, Victoria, 3800, Australia\\
$^{3}$Max-Planck-Institut f\"{u}r Astrophysik, Karl-Schwarzschild-Stra{\ss}e 1, 85748 Garching bei M\"{u}nchen, Germany}

\begin{document}

\maketitle

\begin{abstract} 

Theoretical predictions for the core helium burning phase of stellar evolution are highly sensitive to the uncertain treatment of mixing at convective boundaries.  In the last few years, interest in constraining the uncertain structure of their deep interiors has been renewed by insights from asteroseismology. Recently, Spruit (2015) proposed a limit for the rate of growth of helium-burning convective cores based on the higher buoyancy of material ingested from outside the convective core.  In this paper we test the implications of such a limit for stellar models with a range of initial mass and metallicity.  We find that the constraint on mixing beyond the Schwarzschild boundary has a significant effect on the evolution late in core helium burning, when core breathing pulses occur and the ingestion rate of helium is fastest.  Ordinarily, core breathing pulses prolong the core helium burning lifetime to such an extent that models are at odds with observations of globular cluster populations.  Across a wide range of initial stellar masses ($0.83 \leq M/\text{M}_\odot \leq 5$), applying the Spruit constraint reduces the core helium burning lifetime because core breathing pulses are either avoided or their number and severity reduced.  The constraint suggested by Spruit therefore helps to resolve significant discrepancies between observations and theoretical predictions.  Specifically, we find improved agreement for $R_2$, the observed ratio of asymptotic giant branch to horizontal branch stars in globular clusters; the luminosity difference between these two groups; and in asteroseismology, the mixed-mode period spacing detected in red clump stars in the \textit{Kepler} field.

\end{abstract}


\mathleft

\section{Introduction}
\label{sec:introduction}

In the evolution of low-mass stars, core helium burning (CHeB) is the third stage of nuclear burning, following core- and shell-hydrogen burning, which occur during the main sequence and RGB phases, respectively.  Depending principally on the stellar mass, and therefore their effective temperature, observed CHeB stars may be described as subdwarf B, horizontal branch, RR-Lyrae, red clump, or secondary clump \citep{1999MNRAS.308..818G} stars.  Despite the contrasting surface conditions, the cores of each of these types of stars are similar: they have a central helium-burning convection zone beneath a helium-rich shell where energy transport is dominated by radiation.  Detailed modelling of the evolution of the structure of the cores of these stars poses significant challenges.    Historically, progress in refining models of the mixing in the deep interior of CHeB stars has been impeded by the lack of direct observational constraints or a sufficient physical model. 

\subsection{The importance of the convective boundaries}

The dominant cause of the uncertainty in our understanding of the evolution of CHeB stars is the strong dependence on the treatment of mixing at convective boundaries.  In low-mass (around 1\,M$_\odot$) models, the mass of the convective core does not grow if there is no mixing beyond the Schwarzschild or Ledoux boundary.  However, if there is some mixing beyond the boundary of the convective core and into the radiative region, from e.g. convective overshoot or numerical diffusion, the resulting changes in composition alter the convective stability.  In general, feedback from polluting the radiative zone with the higher opacity products of helium burning, specifically carbon and oxygen, leads to growth in the mass enclosed by the convective core.  Consequently, the CHeB phase lifetime increases substantially -- by around a factor of two \citep[see e.g.][]{2015MNRAS.452..123C,2016MNRAS.456.3866C}.

Depending on the mixing prescription, a partially-mixed, or `semiconvection', zone can develop between the convective core and the convectively stable shell above.  This semiconvection region tends to be marginally stable according to the Schwarzschild criterion and stable according to the Ledoux criterion, because the mass fraction of carbon (or oxygen), and therefore mean molecular weight, decreases outwards.  This configuration can develop either through an explicit numerical scheme \citep[e.g.][]{1971Ap&SS..10..355C,1972ApJ...171..309R,1972ApJ...171..583D} or simply by allowing overshoot at convective boundaries \citep{1986ApJ...311..708L,1990A&A...240..305C}.  In CHeB models, the presence (or not) of semiconvection is important because the semiconvection zone can grow to enclose an amount of mass comparable to the convective core itself, and affect subsequent evolution.

The composition discontinuity that arises at the convective boundary means it is inappropriate to use an extrapolation scheme (e.g. \citealt{2014A&A...569A..63G}) to determine the position of the convective boundary.  Moreover, as burning progresses the magnitude of the discontinuity increases, and the position of the formal convective boundary becomes increasingly unstable to episodes of mixing.  This culminates in core breathing pulses (CBP), the rapid growth in the size of the convective core and ingestion of helium late in CHeB \citep{1973ASSL...36..221S,1985ApJ...296..204C}.  \citet{1973ASSL...36..221S} showed that the convective boundary is unstable to perturbations in the helium abundance because the energy generation rate becomes sensitive to small amounts of mixing when the composition contrast between convective core and its surroundings increases.  Late in CHeB, when the helium abundance is low, even the small absolute changes in helium abundance from overshoot have a large effect on the rates of helium-burning reactions because of the large relative change in the helium abundance.

CBP extend the CHeB lifetime and shorten the subsequent early-asymptotic giant branch phase.  The number, and extent, of CBP that develop is sensitive to numerical treatment and overshooting prescription \citep[e.g.][]{2016MNRAS.456.3866C}.  A number of methods have been employed to suppress CBP, such as omitting the gravitational energy term \citep{1993ApJ...409..387D}, artificially preventing core growth if it would increase the central helium abundance \citep[e.g.][]{1989ApJ...340..241C,1997ApJ...489..822B,2001A&A...366..578C}, and using a non-local scheme for mixing beyond the Schwarzschild boundary \citep[e.g.][]{1986MmSAI..57..411B,2016MNRAS.456.3866C}.

\subsection{Recent progress in modelling}

In this paper we rely primarily on the observational constraints used in the first two papers in this series.  We will not restate these in detail here; instead readers are directed to those papers for background and a summary of the literature.

In \citet{2015MNRAS.452..123C}, hereinafter Paper~I, we compared a suite of stellar models with asteroseismic analysis of core helium burning stars observed during the initial four-year mission of the \textit{Kepler} spacecraft \citep{2012A&A...540A.143M,2014A&A...572L...5M,2016A&A...588A..87V,2017A&A...598A..62M}.  The quantity of particular interest was the inferred asymptotic g-mode period spacing of $\ell = 1$ modes, $\Delta\Pi_1$, because theoretical calculations show its sensitivity to the conditions in and around the convective core.  Crucially, asteroseismology provides the most direct information about the interior structure of CHeB stars.

In \citet{2016MNRAS.456.3866C}, hereinafter Paper~II, we used two constraints derived from the colour-magnitude diagrams of globular clusters: $R_2$, the observed ratio of asymptotic giant branch (AGB) to horizontal branch (HB) stars; and $\Delta \log{L}^\text{AGB}_\text{HB}$, the luminosity difference between the AGB ``clump'' and the horizontal branch.

In both of those studies, the treatment of mixing was the most important uncertainty in the models.  We computed (i) models with a strict Schwarzschild criterion convective boundary, i.e. with neither numerical nor explicit overshooting, and therefore no growth in the mass of the convective core, (ii) models with a region of partial mixing outside the convective core, resulting from either the ``semiconvection'' or ``standard overshoot'' scheme, and (iii) a test case ``maximal overshoot'', which produces the largest possible convective cores.   The ``semiconvection'' sequences were computed by allowing slow mixing in regions formally stable according to the Schwarzschild criterion (i.e. where $\nabla_\text{rad}<\nabla_\text{ad}$).  In those models, the diffusion coefficient $D$ depended on the stability: $\log{D/D_0} \propto \nabla_\text{rad}/\nabla_\text{ad}$ where $D_0$ is a constant, and mixing was only appreciable where $\nabla_\text{rad}$ was close to $\nabla_\text{ad}$.  The standard overshoot models were computed by applying the scheme proposed by \citet{1997A&A...324L..81H} where there is an exponentially decaying diffusion coefficient $D_\text{OS}$ beyond each convective boundary:
\begin{equation}
D_\text{OS}(z) = D_0 e^{-2z/f_\text{OS}H_\text{p}},
\label{eq_herwig}
\end{equation}
where $D_0$ is the diffusion coefficient inside the convective boundary derived from MLT, $z$ is the distance from the boundary, $H_\text{p}$ is the local pressure scale height, and $f_\text{OS}$ is a free parameter.  In the maximal overshoot scheme, the application of convective overshoot was controlled to ensure that the mass of the convective core was as large as possible (i.e. such that $\nabla_\text{rad} > \nabla_\text{ad}$ throughout) and there was no partially mixed region.

Both studies showed that evidence is conclusively against models without mixing beyond the Schwarzschild boundary.  Models with ``standard overshoot'', i.e. overshoot across all convective boundaries, were consistent with both sets of observations, depending on the interpretation of the asteroseismic observations, but only if large CBP late in CHeB are somehow avoided.  Models with maximal overshoot provided a good match to the seismic constraints but were only consistent with the globular cluster observations if the overshooting in the subsequent `gravonuclear' loop phase \citep{1997ApJ...479..279B,1997ApJ...489..822B,2000LIACo..35..529S} was explicitly tuned for the purpose.

Elsewhere, models with a larger overshoot region of $0.5\,H_\text{p}$ have also been tested against asteroseismology.  Regardless of whether the radiative or adiabatic temperature gradient is used in the overshooting region, these models can provide good matches for $\Delta\Pi_1$ in the open clusters in the \textit{Kepler} field \citep{2017arXiv170503077B}.  In that implementation, however, the overshooting is moderated by extending the mixed region $0.5\,H_\text{p}$ from the minimum of $\nabla_\text{rad}$ in the convection zone, which is not necessarily at the boundary \citep{etheses7090}.  Interestingly, this approach, in particular for models with `penetrative convection' where the adiabatic temperature gradient is imposed in the overshooting region, tends to generate a structure very similar to the maximal overshoot models in Papers~I and II \citep[see e.g. Figure 2 and 4 in][]{2015MNRAS.453.2290B}.  

Structures comparable to the maximal overshoot models may also be formed by different means.  \citet{2015ApJ...806..178S} applied atomic diffusion without overshoot to yield monotonic growth in the mass of the convective core (which similarly occurs due to numerical diffusion in {\sc monstar} models without explicit overshoot and an insufficiently resolved mesh; see e.g. the upper panel of Figure~1 in \citealt{2016AN....337..788C}).  \citet{2016ApJ...827....2V} and \citet{2017arXiv170605454D} produce a similar structure by choosing appropriate parameters for the mixing scheme described by Equation~\ref{eq_herwig}: they use a small value for the parameter $f_\text{OS}$, which is then reduced further late in CHeB; they also measure $D_0$ from close to the convective boundary and replace it with a fraction of the thermal diffusivity $K$ when $D_0 < K$.

Recently, \citet{2015A&A...582L...2S} deduced a limit on the growth of helium-burning convective cores.  This constraint exists because where there is a difference between the mean molecular weight of the material on either side of a convective boundary, the growth of the convection zone from ``overshooting'' should not exceed a rate that would imply that ascending fluid elements are less buoyant than descending fluid elements.  In the case of a helium-burning convective core expanding in mass, the cooler fluid elements descending from the outer boundary of the convection zone are made less dense, and more buoyant, by the entrainment of higher-helium material from beyond the boundary.  

In preliminary calculations, \citet{2015A&A...582L...2S} found that the maximum rate of growth predicted from the structure of CHeB stellar models is of the same order of magnitude as the core growth rate inferred from observations of (approximately solar mass) CHeB stars.  This finding strongly suggests that this new limit may be an important consideration for theoretical predictions of the evolution of the CHeB phase.  In this paper we test the implications of applying this core growth constraint throughout the evolution of full stellar models.

\section{Implementation of the mixing scheme}
\label{sec:method}

We determine the maximum entrainment rate using the same approximation as \citet{2015A&A...582L...2S}, namely that the core is a fully ionized helium-carbon mixture.  We therefore arrive at an expression, equivalent to Equations~12 and 16 in \citet{2015A&A...582L...2S}, for the maximum rate at which helium may be ingested:
\begin{equation}
\dot{m}_{i}=\alpha_{i} \frac{12}{5} \frac{L}{RT} \left ( 1- \frac{\nabla_\text{ad}}{\nabla_\text{rad}} \right),
\label{eq_mmix}
\end{equation}
where $\alpha_{i}$ is the ingestion efficiency, $R$ is the gas constant, and $L$, $T$, $\nabla_\text{rad}$, and $\nabla_\text{ad}$ are the luminosity, temperature, radiative temperature gradient, and adiabatic temperature gradient, respectively, at the inside boundary of the convection zone.  We shall hereinafter refer to this scheme as ``Spruit overshoot'' (SOS). 

We apply SOS after the ignition of helium and only for convective boundaries in the hydrogen-free core.  After each time step $\Delta t$, Equation~\ref{eq_mmix} is used to determine the extent of mixing, i.e. $\Delta m = \dot{m}_i \Delta t$.  At outer convective boundaries (where the helium abundance is typically higher outside the boundary than it is inside the convection zone) the procedure is as follows:
\begin{enumerate}[leftmargin=*]
\item Loop outward over the mesh beginning at the first radiative zone $j_1$.
\item At each step in the loop, zone $j_n$ say, calculate the mass of helium $\Delta m$ that would be transported into the convective region if the convection zone were to be fully mixed with the material in zones $j_1$ to $j_n$.  This is calculated from 
\begin{equation}
\label{eq:loop}
\Delta m = \dot{m}_i \Delta t = \sum_{j=j_1}^{j_n} (Y_j-Y_{j_0}) \Delta m_j,
\end{equation}
where $\Delta m_j$ is the mass in each shell, $Y_j$ is the helium mass fraction, and $Y_{j_0}$ is the helium mass fraction in the convection zone.
\item Continue in the loop until the largest $j_n$ is found for which the rate of helium ingestion $\dot{m}_i$ is still less than the limit from Equation~\ref{eq_mmix}.
\item Homogenize the chemical species in the region that includes the convection zone and the radiative zone up to $j_n$.
\end{enumerate}
At inner convective boundaries (which arise when $\nabla_\text{rad}/\nabla_\text{ad}$ reduces and a radiative zone appears inside a convective region) we also calculate the maximum rate helium can be mixed across the new Schwarzschild boundary.  In that case the radiative region is likely to be less helium-rich than the convective region, i.e. $Y_j-Y_{j_0} < 0$, so the same procedure is applied except the loop proceeds in the opposite direction and $Y_{j_0} - Y_j$ is used instead of $Y_j-Y_{j_0}$ in Equation~\ref{eq:loop}.  We use a very fine mesh near convective boundaries, typically of the order $\Delta m_j \approx 10^{-8}\,\text{M}_\odot$, to ensure that the rate at which helium actually mixes, $\dot{m}_i$ in Equation~\ref{eq:loop}, closely matches the calculated limit $\dot{m}_i$ in Equation~\ref{eq_mmix}, i.e. $\Delta m_j \ll \Delta m$.

In our models, mixing proceeds precisely as far as specified by the algorithm, i.e. we always apply the \textit{maximum} rate of ingestion indicated by Equation~\ref{eq_mmix} (after including the efficiency factor).  In the evolution sequences computed for this paper, the maximum time step ($\Delta t = 2.5 \times 10^3\,\text{yr}$ for the models with $M \leq 1\,\text{M}_\odot$ for example) is short enough for $\nabla_\text{rad}/\nabla_\text{ad}$ at the boundary to not change significantly between time steps.  We also reduce the time step when necessary in order to limit the distance over which overshoot mixing can occur to at most 1.0\,$H_\text{p}$.  This is generally not important, however, because the limit is most likely to be reached if the composition is nearly uniform, and in that case, with a well resolved mesh, $1- \nabla_\text{ad}/\nabla_\text{rad}$ vanishes near the convective boundary (and so too, therefore, does the rate of ingestion).

In our implementation, the extent of overshoot is determined only by the conditions at the convective boundary.  Apart from the mesh point immediately adjacent to the boundary, the conditions inside convection zones are not used to limit the rate of mixing.  This is the case even if $\nabla_\text{rad}/\nabla_\text{ad}$ is smaller inside the convection zone than at the boundary, and which if used in Equation~\ref{eq_mmix} would imply a lower ingestion rate.  Although we do not investigate it in this paper, there is an interesting similarity between the SOS scheme where the minimum $\nabla_\text{rad}/\nabla_\text{ad}$ anywhere in the convection zone in Equation~\ref{eq_mmix}, and the models with maximal overshoot in Paper~I and Paper~II.  In both cases, the rate that helium is mixed into the core decreases to zero as the minimum $\nabla_\text{rad}/\nabla_\text{ad}$ approaches unity (usually at a coordinate deep inside the convective core).

It is also worth pointing out the similarities and differences between SOS and other prescriptions in the literature.   In a number of other schemes, the extent of overshooting also depends on the value of $\nabla_\text{rad}/\nabla_\text{ad}$ at the convective boundary.  \citet{1971Ap&SS..10..340C} arrived at a ``velocity'' of convective core growth: 
\begin{equation}
v = 10^{-5} \frac{1-\nabla_\text{ad}/\nabla_\text{rad}}{1-\mu^e / \mu^i},
\label{eq_castellani_ingestion}
\end{equation}
where $\mu^i$ and $\mu^e$ denote the mean molecular weight on the interior and exterior sides of the convective core boundary, respectively.
Similarly, \citet{1991ASPC...13..299S} used
\begin{equation}
v = F_\text{ov} \frac{\nabla_\text{rad}-\nabla_\text{ad}}{\mu^i - \mu^e},
\end{equation}
where $F_\text{ov}$ is an uncertain parameter (and was set to $10^{-5}$ in tests by \citealt{1991ASPC...13..299S}, nearly identical to Equation~\ref{eq_castellani_ingestion}).  Dependence on $\nabla_\text{rad} / \nabla_\text{ad}$ is also indirectly present in the \citet{1997A&A...324L..81H} scheme (Equation~\ref{eq_herwig}), where there is an exponential decay in the diffusion coefficient, and which is used for the standard overshoot models in this paper.  In that formulation, the extent of overshoot diminishes as $\nabla_\text{rad}$ approaches $\nabla_\text{ad}$ near the edge of the convection zone, via the MLT prediction for convective velocity.

In the SOS scheme and the three other examples, the rate of ingestion across the convective boundary is (theoretically at least) time-step independent.  This contrasts with, e.g., mixing a fixed distance, such as a fraction of a pressure-scale-height, at each time step.  In the CHeB case, the importance of this distinction for the evolution is amplified by the fact that the position of the convective boundary moves in response to mixing episodes.

\section{Stellar models}
\label{sec:stellar_models}

In this study we compute stellar models with the Monash University stellar structure code {\sc monstar}, which has been described previously \citep[e.g.][ and references therein]{2008A&A...490..769C,2014ApJ...784...56C}.  We choose some representative stellar models: a 1\,M$_\odot$ solar-metallicity model that corresponds to those in Paper~I, a $M = 0.83\,\text{M}_\odot$ and $\text{[Fe/H]}=-1$ model that corresponds to the globular cluster models in Paper~II, and some higher-mass solar-metallicity models ($2$, $5$, and $10\,\text{M}_\odot$).  The globular cluster model has initial helium content $Y=0.245$, red giant branch mass loss rate given by \citet[][with $\eta = 0.4$]{1975MSRSL...8..369R}, and the mass at the beginning of CHeB is 0.67\,M$_\odot$. The other models have $Y = 0.278$ and no mass loss.  Both sets of models have a solar calibrated MLT parameter $\alpha_\text{MLT} = 1.60$.

The low-mass ($M \leq 2\,\text{M}_\odot$) solar-metallicity and globular cluster models are suitable for comparisons with the key constraints from asteroseismology and observations of globular cluster stars, respectively.  Such a modest range of model parameters is sufficient to explore the effects of SOS because compared with the uncertainties regarding mixing, CHeB evolution is relatively insensitive to the initial composition and other uncertainties (Paper~II).  

We analyse models with a wide range of `ingestion efficiency', $\alpha_i$ in Equation~\ref{eq_mmix}.  This includes the models within the physically consistent range, $0.0 \leq \alpha_i \leq 1.0$, and a handful of models with $\alpha_i > 1.0$ as an extreme test of the sensitivity.  The models with `standard overshoot' use the mixing scheme given in Equation~\ref{eq_herwig}, with $f_\text{OS} = 0.001$, consistent with Papers~I and II.

\section{Evolution with Spruit overshoot}

\begin{figure}
\includegraphics[width=\linewidth]{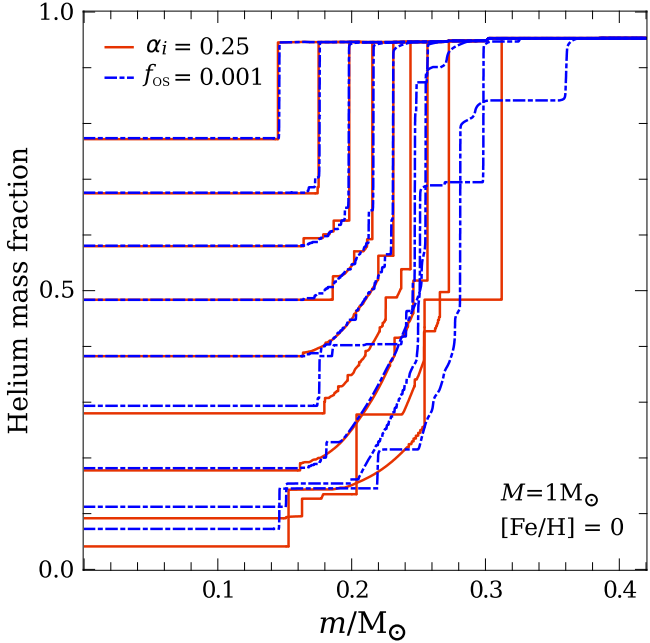}
  \caption{Evolution of internal helium mass fraction during core helium burning for 1\,M$_\odot$ solar-metallicity models with standard overshoot with $f_\text{OS} = 0.001$ (blue dash-dot lines) and SOS with $\alpha_{i} = 0.25$ (solid red lines).  Lines are plotted at 12.5\,Myr intervals.}
  \label{figure_internal_helium_evolution}
\end{figure}

Figure~\ref{figure_internal_helium_evolution} shows the evolution of the internal helium abundance for a model with standard overshoot and another with SOS and $\alpha_{i} = 0.25$, half the `reasonable upper limit' suggested by \citet{2015A&A...582L...2S}.  The evolution of the central helium abundance and total mass of the convective core for these two cases is nearly identical for the first half of CHeB, until $Y_\text{cent} \approx 0.4$.  After this, small differences occur, but the structures are again almost identical when $Y_\text{cent} \approx 0.18$.  A lasting divergence between the two sequences only emerges very late in CHeB.  The scale of this difference is not unexpected given the extent of the contrast between standard overshoot sequences computed with different parameters (see e.g. Figure~12 in Paper~II).

\begin{figure}
\includegraphics[width=\linewidth]{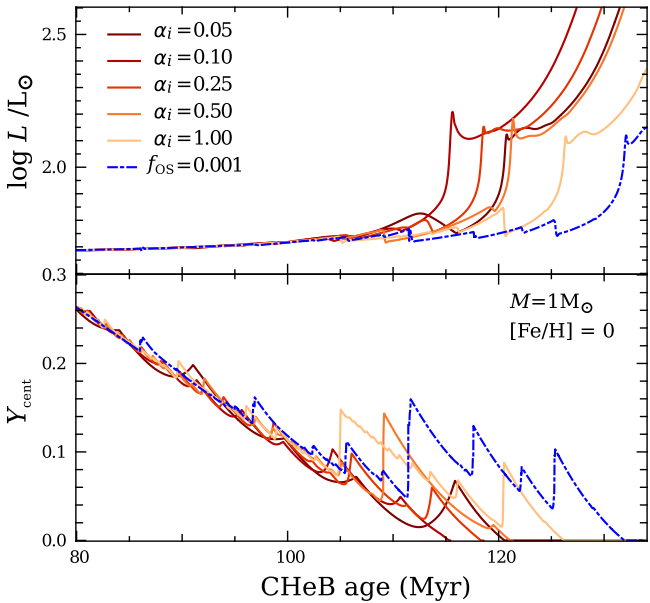}
  \caption{Evolution of the surface luminosity (upper panel) and central helium mass fraction (lower panel) for 1\,M$_\odot$ solar-metallicity models near the conclusion of CHeB.  The models are calculated using standard overshoot with $f_\text{OS} = 0.001$ (blue dash-dot line) and SOS with $\alpha_{i} = 0.05$, $0.10$, $0.25$, $0.50$, and $1.0$ (denoted by solid red lines of increasing brightness).}
  \label{figure_central_helium_evolution}
\end{figure}

Figure~\ref{figure_central_helium_evolution} shows the evolution of the surface luminosity and central helium mass fraction for 1\,M$_\odot$ SOS models with a range of ingestion efficiency $0.05 \leq \alpha_i \leq 1.0$ and for a standard overshoot model.  The evolution of the models is similar for most of the CHeB phase: substantial differences in the luminosity and the central helium abundance are only apparent after more than about 100\,Myr.  During the earlier quiescent period, the position of the convective core boundary is relatively stable and the models typically experience only small increases in the central helium abundance of $\Delta Y_\text{cent} \lesssim 0.02$ and the overall central helium depletion is consistent between all of the models.  Similarly, differences in the helium burning luminosity are small: typically we find $\Delta \log{L_\text{Heb}/\text{L}_\odot} \lesssim 0.02$ until about 100\,Myr, when $Y_\text{cent} \approx 0.1$.

Despite this similarity, and the essentially identical surface luminosity, a difference is evident: the sequences with lower $\alpha_i$ experience slower episodes of helium ingestion into the core.  In the models with $\alpha_i \leq 0.1$ these generally take at least 1\,Myr, compared with around 0.1\,Myr in standard overshoot sequences.  In contrast, in the models with higher ingestion efficiency ($\alpha_i$ = 0.5, 1.0) helium is ingested at a rate comparable with that of the standard overshoot sequences.  Models with standard overshoot and SOS with $0.05 \leq \alpha_i \leq 1.0$ can have matching helium abundance profiles until fairly late into the CHeB phase (such as the two models with $Y_\text{cent} \approx 0.18$ in Figure~\ref{figure_internal_helium_evolution}).  This is possible because the \textit{average} rates of both helium ingestion and burning are generally consistent between these models until near the end of CHeB, when CBP occur.

The most obvious divergence appears late in CHeB when $ Y_\text{cent} \lesssim 0.1$.  The SOS models experience fewer, and smaller, increases in $Y_\text{cent}$.  The standard overshoot sequence includes several CBP where the central helium is replenished whereas the SOS sequences experience no more than two CBP with $\Delta Y_\text{cent} > 0.04$ and consequently have a significantly curtailed CHeB lifetime.  In Section~\ref{sec_cbp} we examine the mechanism by which the SOS models can avoid CBP.  The extent to which SOS reduces the CHeB lifetime does not show a straightforward dependence on $\alpha_i$ (in the range of $\alpha_i$ tested here; see Section~\ref{section_gc} for tests of the effect of a wider range of $\alpha_i$).  There is a spread in CHeB lifetime of more than 10\,Myr which is comparable to that seen for standard overshoot models computed with different parameters (see Paper~II).

When SOS is used, the significantly moderated ingestion towards the end of CHeB reduces the availability of helium for the $^{12}\text{C}(\alpha,\gamma)^{16}\text{O}$ reaction.  Consequently, the evolution sequences with SOS finish CHeB with a slightly higher C/O ratio in the degenerate core.  The final difference in carbon mass fraction is around generally $\Delta X_\text{C} = 0.01$, but up to $\Delta X_\text{C} = 0.08$ for the $\alpha_i = 0.1$ model, which is the shortest-lived.

\subsection{Comparison with asteroseismology}

In the new field of red clump asteroseismology, the key constraint for models is the period spacing between $\ell = 1$ g-modes.  Observed oscillations of mixed g-mode and p-mode character \citep[e.g,][]{2010ApJ...713L.176B} may be analysed to infer the asymptotic limit of the $\ell = 1$ g-mode period spacing \citep[e.g.][]{2012A&A...540A.143M,2014A&A...572L...5M,2016A&A...588A..87V}, which is known theoretically:
\begin{equation}
\Delta\Pi_1 = \frac{2\pi^2}{\sqrt{2}}  \left[ \int\limits_\text{}^{}{\frac{N}{r}\text{d}r} \right]^{-1},
\end{equation}
where $N$ is the Brunt--V{\"a}is{\"a}l{\"a} frequency, and the integral is over the region with $N^2 > 0$ \citep{1977AcA....27...95D}.  However, here we shall generically refer to this observationally derived period spacing as $\Delta P_1$ to account for the possibility that it differs from the theoretical asymptotic value.  In Paper~I we showed that standard overshoot models can match the asteroseismology, but only if, as suggested by our pulsation calculations, mode trapping at the outer boundary of the partially mixed zone increases the period spacing between the most observable modes, and therefore $\Delta P_1$.  Whereas $\Delta\Pi_1$ depends on the radial extent of the convective core \citep{2013ApJ...766..118M}, if the mode-trapping scenario proposed in Paper~I is correct then $\Delta P_1$ is actually dependent on the radial extent of the partially mixed zone.  This is because certain modes are trapped by the sharp composition profile at the outer boundary of the partially-mixed region, while the remainder are insensitive to the partially mixed region and behave as though it is not part of the g-mode cavity.  We adopt this assumption in order to compute $\Delta P_1$, which we then compare with observations.  Supporting this assumption, pulsation calculations for main sequence models have shown this effect \citep{2008MNRAS.386.1487M}, and more recently \citet{2017MNRAS.465.1518G} have reported in depth the effects of mode-trapping in CHeB models using independent stellar evolution and pulsation codes.  

The similarity between models with Spruit and standard overshoot in both convective core size and the partially mixed region size is shown in Figure~\ref{figure_internal_helium_evolution}.  Irrespective of mode trapping, predictions for $\Delta P_1$ for most of the evolution are therefore unaffected by adopting the constraint from \cite{2015A&A...582L...2S}.  The differences are only substantial later in the evolution, coinciding with the occurrence of CBP when the evolution is most susceptible to small differences in the extent of overshoot and the numerical treatment (see Figures~14 and 15 in Paper~II).  

\begin{figure}
\includegraphics[width=\linewidth]{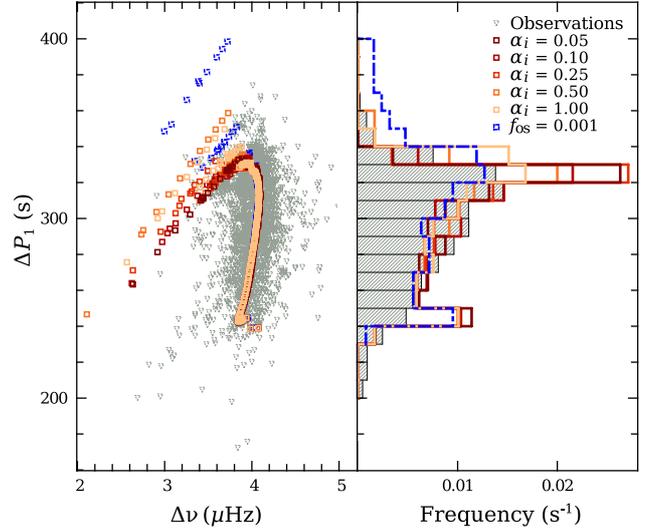}
  \caption{Comparison of g-mode period spacing between models and observations.  \textbf{Left panel:} The evolution of 1\,M$_\odot$ models computed using standard overshoot (blue dash-dot squares) and SOS with $\alpha_i = 0.05$, $0.10$, $0.25$, $0.50$, and $1.0$ (red squares of increasing brightness) in $\Delta\nu - \Delta P_1$ space plotted at 1\,Myr intervals.  The inferred asymptotic g-mode period spacings $\Delta\Pi_1$ for stars in the \textit{Kepler} field with asteroseismic masses $0.8 < M/\text{M}_\odot < 1.25$ from \citet{2016A&A...588A..87V} are denoted by dotted grey triangles.  \textbf{Right panel:} Histogram of the frequency distribution of $\Delta P_1$ for observations (grey hatched bars) and predictions from models (same colours as left panel) when $\Delta \nu > 2.5\,\mu$Hz.  Here $\Delta P_1$ is the predicted $\ell = 1$ g-mode period spacing for the most observable modes, which is calculated in the same way as $\Delta\Pi_1$ except the region inside the outermost composition discontinuity in the partially mixed region is excluded, in accordance with the mode-trapping scenario proposed in Paper~I.}
  \label{figure_DP1}
\end{figure}

Within the mode-trapping scenario, we would expect the reduction in the maximum extent of the partially mixed region to reduce the peak $\Delta P_1$.  In Figure~\ref{figure_DP1} we present a comparison between observations and predictions for $\Delta P_1$ (accounting for mode trapping in the same way as in Paper~I) for 1\,M$_\odot$ models with SOS ($\alpha_i = 0.05$, $0.10$, $0.25$, $0.5$, and $1.0$) and with standard overshoot (with $f_\text{OS} = 0.001$).  These are plotted in the $\Delta\nu-\Delta P_1$ plane, where $\Delta\nu$ is the mean large frequency separation (which is a measure of the sound crossing time of the star).  In this space, models evolve from low to high $\Delta P_1$ at nearly constant $\Delta \nu$ for the bulk of the CHeB phase, then to lower $\Delta\nu$ very late in CHeB.  The observations are \textit{Kepler} field stars with asteroseismic-determined mass $0.8 < M/\text{M}_\odot < 1.25$ from Figure~1 in \citet{2016A&A...588A..87V}.

In Table~\ref{table_partially_mixed_extent} we show the maximum mass enclosed by the partially mixed region outside the convective core (which invariably happens at the end of CHeB).  This is strongly dependent on the late-CHeB evolution and as a proxy for radius gives an indication of the peak $\Delta P_1$ expected within the mode-trapping scenario.  SOS models with lower $\alpha_i$ tend to have a less massive partially mixed region (although the dependence upon $\alpha_i$ is not monotonic) and consequently the sequences with lower $\alpha_i$ generally do not predict more stars with high $\Delta P_1$ than are observed (Figure~\ref{figure_DP1}).  Increasing $\alpha_i$ only has a small effect on the mass enclosed by the partially mixed region because if $\alpha_i$ is sufficiently large ($\alpha_i \gtrsim 0.05$ in our tests) the outermost convection zone in the core will advance until it is only marginally unstable ($\nabla_\text{rad}/\nabla_\text{ad} \approx 1$), when the ingestion rate vanishes (Equation~\ref{eq_mmix}).  An example of this can be seen at $m = 0.28\,\text{M}_\odot$ in the model with SOS and $\alpha_i = 0.25$ shown in Figure~\ref{figure_CBP_ratgrads} (which is discussed in Section~\ref{sec_cbp}).

In all of the Spruit overshoot models shown in Figure~\ref{figure_DP1} the mixing during the bulk of CHeB is relatively slow compared with other schemes such as the $f_\text{OS} = 0.001$ sequence, or those with instantaneous overshoot, e.g., those from \citet{2013ApJ...766..118M}.   The choice of $\alpha_i$ in that range ($0.05 \leq \alpha_i \leq 1.0$) has only a small effect on the radius of the edge of the partially mixed zone, and therefore $\Delta P_1$.  More substantial differences occur later in CHeB but this does not have a large effect on the typical $\Delta P_1$ for a number of reasons: i) most of the $\Delta P_1$ evolution has already been determined by this time; and ii) the convective core and partially mixed zone are more C- and O-rich, and therefore more dense, reducing the radial extent of the partially mixed zone.  When $\alpha_i$ is small, the ingestion rate, and therefore the growth of the partially mixed core, is always limited by the condition from \citep{2015A&A...582L...2S}, so the extent of overshoot is typically much less than is the case with instantaneous overshoot, for example.

The standard overshoot result is consistent with the two equivalent example sequences in Paper~I that predict a maximum $\Delta P_1$ slightly higher than is observed.  The highest $\Delta P_1$ during the evolution is attained following CBP, when the partially mixed region extends furthest.  The speed of CBP is responsible for the isolated sequences at high-$\Delta P_1$ in the left-hand panel of Figure~\ref{figure_DP1}, which are most pronounced with standard overshoot.  In contrast, the models with SOS undergo fewer CBP and therefore the peak values of $\Delta P_1$ and the predicted number of stars with high $\Delta P_1$ (larger than $340\,\text{s}$) are both reduced, better conforming with observations.  The difference in predictions for $\Delta P_1$ among these models is almost entirely due to the extent of CBP: sequences with more CBP tend to have a longer lifetime and finish the CHeB phase with a larger partially mixed region, increasing $\Delta P_1$.  Indeed, were another method employed to suppress CBP in the standard overshoot sequence, its $\Delta P_1$ evolution would closely match that for SOS models, and similarly better match the observations.

Each of the distributions for SOS sequences in Figure~\ref{figure_DP1} show peaks at low- and high-$\Delta P_1$ corresponding to the beginning and later part of CHeB, respectively.  The absence of such peaks in the observed data could be explained by the diversity in the \textit{Kepler} field: a variation of helium and metallicity of $\Delta Y = 0.1$ and $\Delta \text{[Fe/H]} = 0.05$ could each explain a spread of $\Delta P_1$ of greater than 10\,s \citep[see e.g.][]{2017arXiv170503077B} and this would be exacerbated after accounting for the stochastic evolution of the models in this paper.  Additionally, observational and fitting errors would tend to smooth the peaks in the $\Delta P_1$ distribution.  We note that taking advantage of the homogeneous populations in the open clusters in the \textit{Kepler} field can significantly reduce the uncertainty from the stellar properties, however, this also introduces a new challenge resulting from their relatively small CHeB populations \citep{2017arXiv170503077B}.

Our models with a solar-calibrated MLT parameter $\alpha_\text{MLT} = 1.60$ automatically match $\Delta\nu$ during the CHeB phase.  However, prior to the cessation of core convection they rapidly cross the $\Delta\nu-\Delta P_1$ diagram (which is also seen for the various CHeB models in Figure~3 in \citealt{2015MNRAS.453.2290B}) and therefore predict the existence of more stars around $\Delta\nu \approx 3\,\mu\text{Hz}$ and $\Delta P_1 \approx 300\,\text{s}$ than are observed.  It is possible that a selection effect is responsible for the absence of observed stars in this part of the diagram.  Alternatively, if $\alpha_\text{MLT}$ were increased by around $0.5$, the models would have higher $\Delta\nu$ near the end of CHeB and they would match the stars with $\Delta\nu \approx 3.4\,\mu\text{Hz}$.  Stellar properties from the APOKASC catalogue \citep{2014ApJS..215...19P}, $T_\text{eff}$ and $\log{g}$, suggest these are indeed CHeB stars.  If $\alpha_\text{MLT}$ is assumed to be constant throughout the evolution, however, the agreement for the bulk of CHeB stars (with $\Delta\nu \approx 4\,\mu\text{Hz}$) would be worsened.

\setlength{\tabcolsep}{4pt}
\begin{table}
\begin{center}
  \caption{Maximum mass (in units of M$_\odot$) enclosed by the partially mixed region (i.e. the combined mass of the convective core and the partially mixed region surrounding it) during the CHeB evolution for low-mass models with different SOS ingestion efficiency and with standard overshoot.}
  \label{table_partially_mixed_extent}
 \footnotesize
\begin{tabular}{cccc}
\hline
  &  `Globular cluster' & `Solar-like' \\
  & $M=0.67\,\text{M}_\odot$ & $M=1\,\text{M}_\odot$  \\
 $\alpha_i$ & $\text{[Fe/H]} = -1.0$ & $\text{[Fe/H]} = 0.0$ \\
\hline
0.05 & 0.293 & 0.278 \\
0.10 & 0.309 & 0.280 \\
0.25 & 0.308 & 0.290 \\
0.50 & 0.338 & 0.312 \\
1.00 & 0.321 & 0.306 \\
\hline
$f_\text{OS} = 0.001$ & 0.346 & 0.360 \\
\hline
\end{tabular}
\normalsize
\end{center}
\end{table}

\subsection{Comparison with globular cluster star counts}
\label{section_gc}

In Paper~II we determined empirically $R_2$, the observed ratio of asymptotic giant branch to horizontal branch stars in globular clusters, and $\Delta\log L_\text{HB}^\text{AGB}$, the luminosity difference between these two groups.  We found that $R_2 = 0.117 \pm 0.005$ and $\Delta \log{L}_\text{HB}^\text{AGB} = 0.455 \pm 0.012$ for Galactic globular clusters without a blue extension to the horizontal branch.  The evolution of models with standard overshoot, while chaotically dependent on numerical treatment and the overshooting formulation, generally predicts $R_2$ significantly lower than is observed (around $R_2 \lesssim 0.10$).  While the predictions for $\Delta \log{L}_\text{HB}^\text{AGB}$ are consistent on average, the stochastic occurrence of CBP produces a wider AGB clump in the luminosity probability density function compared with globular cluster populations and this discrepancy is further worsened when a population of models is considered.  

\begin{figure}
\includegraphics[width=\linewidth]{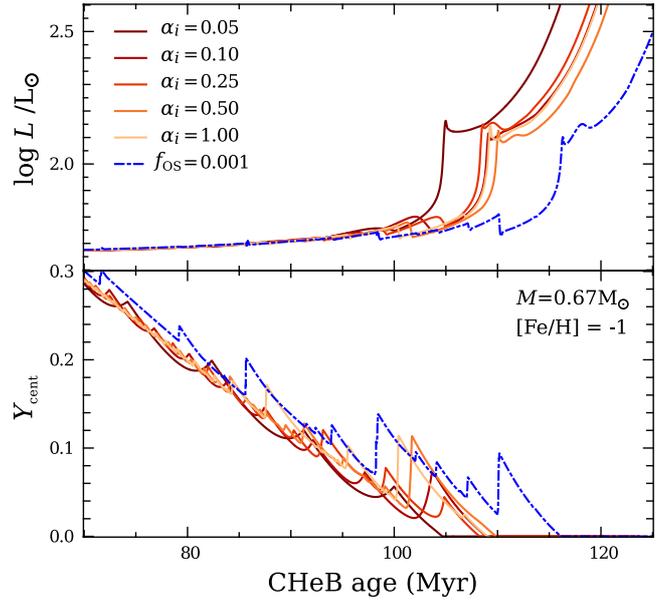}
  \caption{Surface luminosity (upper panel) and central helium abundance (lower panel) evolution for the representative globular cluster models near the end of CHeB and during the early-AGB.  The models are calculated using standard overshoot with $f_\text{OS} = 0.001$ (blue dash-dot line) and SOS with ingestion efficiency $\alpha_{i} = 0.05$, $0.10$, $0.25$, $0.50$, and $1.0$ (solid red lines of increasing brightness).}
  \label{figure_gc_L_ev}
\end{figure}

In Table~\ref{table_gc} we summarize the evolution of our suite of `globular cluster' models with different mixing prescriptions.  We compare SOS sequences computed for a range of $\alpha_{i}$ with a standard-overshoot sequence.  We choose reasonable estimates of the ingestion efficiency, $\alpha_i \leq 1.0$, as well as some higher values, $\alpha_i > 1.0$ (and higher than is consistent with the theory).  In the sequences with $0.05 \leq \alpha_i \leq 1.0$ the CHeB lifetime is reduced compared with the standard overshoot by an average of 6\,Myr (where the average is computed from a suite of models with increments of $\Delta \alpha_i = 0.05$).  The evolution of the surface luminosity and central helium abundance for a sample of these models is shown in Figure~\ref{figure_gc_L_ev}.

The SOS models with $0.05 \leq \alpha_i \leq 1.0$ predict $R_2 = 0.101 \pm 0.007$ (where the uncertainty is one standard deviation), which is closer to observed value of $R_2 = 0.117 \pm 0.005$ than implied from standard overshoot models ($R_2 = 0.090$ in this example and typically $R_2< 0.10$; see Paper~II).  Similarly, the luminosity difference between the HB and the AGB $\Delta \log{L}_\text{HB}^\text{AGB} = 0.449 \pm 0.025$ is consistent with observed value of $\Delta \log{L}_\text{HB}^\text{AGB} = 0.455 \pm 0.012$.  Moreover, excluding the sequences with $\alpha_i > 0.50$ (the `reasonable upper limit' suggested by \citealt{2015A&A...582L...2S}) improves the agreement with observations: for these we find $R_2 = 0.105 \pm 0.005$ and $\Delta \log{L}_\text{HB}^\text{AGB} = 0.449 \pm 0.015$.  We do not find a monotonic dependence of $R_2$, $\Delta \log{L}_\text{HB}^\text{AGB}$, and the CHeB lifetime on $\alpha_i$ when $0.05 \leq \alpha_i \leq 1.0$.  Instead, we observe a slight trend of increasing CHeB lifetime and decreasing $R_2$ with an increase in $\alpha_i$ beneath a significant scatter, which is driven by the chaotic evolution late in CHeB, when CBP occur.

Throughout the part of the evolution shown in Figure~\ref{figure_gc_L_ev} the position of the convective core boundary is most unstable for the standard overshoot model, and consequently the central helium abundance is more variable (similar to the solar-mass and solar-metallicity models in Figure~\ref{figure_central_helium_evolution}).  In this particular example, much of the longer lifetime for the standard overshoot model results from the breathing pulse that occurs after 110\,Myr.  Prior to about 5\,Myr before that event, the luminosity and central helium abundance evolution is similar to most of the SOS models.  The difference between $R_2$ for the standard overshoot sequence and the SOS models with $\alpha \leq 1.0$ would be minimized if CBP were avoided after 100\,Myr.

In contrast with the evolution sequences with moderate ingestion efficiency, those with very low ingestion efficiency ($\alpha_i \leq 0.025$) provide a worse match to the observed $R_2$ than standard overshoot models.  In the model with $\alpha_i = 0.01$, the CHeB lifetime is even longer than the standard overshoot sequence and the agreement with observations is poor: $R_2 = 0.060$, $\Delta \log{L}_\text{HB}^\text{AGB} = 0.641$.  In that sequence, the slower expansion of the convection zone prevents it from dividing into two until after more than 70\,Myr.  This split occurs only once.  By the end of CHeB there is no partially mixed region and the mass enclosed by the convection zone is 0.30\,M$_\odot$, much larger than for the higher $\alpha_i$ models.  Interestingly, the absence of a partially mixed region implies a better match to the high $\Delta\Pi_1$ inferred for stars in the \textit{Kepler} field, without invoking mode trapping.  However, this structure is ruled out by the inability to match $R_2$ and $\Delta \log{L}_\text{HB}^\text{AGB}$.  Lastly, we note that the overall evolution of the $\alpha_i = 0.01$ sequence is remarkably similar to those with `maximal overshoot' explored in Papers~I and II.

At the opposite end of the spectrum, when $\alpha_i$ is large, the CHeB lifetime is also increased (to an average of 121\,Myr for models with $\alpha_i > 1.0$ compared with 111\,Myr for models with $0.05 \leq \alpha_i \leq 1.0$) and $R_2$ reduced relative to the sequence with standard overshoot ($0.085$ compared with $0.101$).  In the SOS sequences with high $\alpha_i$, the mass enclosed by the convective core remains relatively stable once a partially mixed region is established (generally fluctuating by less than 0.02\,M$_\odot$).  Late in CHeB, however, the faster ingestion rate is more conducive to CBP, and as a result more helium is mixed into the convection zone, extending the lifetime and reducing $R_2$.  In the next section we analyse how the development of CBP depends on the helium ingestion rate.

\setlength{\tabcolsep}{2pt}
\begin{table}
\begin{center}
  \caption{Summary of the CHeB evolution for models representative of globular cluster stars with different SOS ingestion efficiency $\alpha_i$.  Model properties are described in Section~\ref{section_gc}.}
  \label{table_gc}
 \footnotesize
\begin{tabular}{cccc}
\hline
 $\alpha_i$ &  $R_2$ & $\Delta \log{L}_\text{HB}^\text{AGB}$ & $t_\text{HB}$ (Myr) \\
\hline
0.010 & 0.060 & 0.641 & 120.4 \\
0.025 & 0.089 & 0.497 & 112.2 \\
0.050 & 0.114 & 0.457 & 106.2 \\
0.075 & 0.104 & 0.465 & 109.3 \\
0.100 & 0.103 & 0.457 & 110.4 \\
0.125 & 0.109 & 0.461 & 107.8 \\
0.150 & 0.109 & 0.465 & 107.7 \\
0.175 & 0.108 & 0.465 & 108.4 \\
0.200 & 0.105 & 0.437 & 110.2 \\
0.225 & 0.108 & 0.457 & 108.1 \\
0.250 & 0.103 & 0.469 & 109.8 \\
0.275 & 0.107 & 0.461 & 108.4 \\
0.300 & 0.100 & 0.421 & 112.6 \\
0.325 & 0.101 & 0.433 & 112.2 \\
0.350 & 0.106 & 0.445 & 109.8 \\
0.375 & 0.095 & 0.445 & 113.8 \\
0.400 & 0.107 & 0.445 & 108.9 \\
0.425 & 0.108 & 0.437 & 108.5 \\
0.450 & 0.096 & 0.457 & 112.9 \\
0.475 & 0.100 & 0.449 & 112.2 \\
0.500 & 0.103 & 0.433 & 111.4 \\
0.525 & 0.103 & 0.445 & 110.7 \\
0.550 & 0.098 & 0.465 & 112.9 \\
0.600 & 0.100 & 0.473 & 111.3 \\
0.650 & 0.096 & 0.469 & 113.9 \\
0.700 & 0.111 & 0.433 & 107.9 \\
0.750 & 0.105 & 0.449 & 109.6 \\
0.800 & 0.096 & 0.481 & 112.4 \\
0.850 & 0.100 & 0.461 & 112.1 \\
0.900 & 0.095 & 0.369 & 115.4 \\
0.950 & 0.080 & 0.425 & 119.5 \\
1.00  & 0.102 & 0.453 & 110.6 \\
1.25  & 0.089 & 0.353 & 119.1 \\
1.50  & 0.075 & 0.381 & 126.3 \\
1.75  & 0.096 & 0.413 & 112.9 \\
2.00  & 0.102 & 0.457 & 111.1 \\
4.00  & 0.089 & 0.365 & 117.9 \\
5.00  & 0.092 & 0.469 & 114.2 \\
7.50  & 0.091 & 0.425 & 116.4 \\
10.0  & 0.076 & 0.397 & 125.7 \\
15.0  & 0.097 & 0.461 & 116.3 \\
20.0  & 0.082 & 0.477 & 138.3 \\
50.0  & 0.085 & 0.477 & 122.0 \\
100   & 0.077 & 0.469 & 116.7 \\
1000  & 0.054 & 0.473 & 130.1 \\
\hline
 Average & & & \\
\scalebox{0.9}[0.9]{$0.05 \leq \alpha_i \leq 1$} & 0.101$ \pm $0.007 & 0.449$ \pm $0.025 & 111.3$ \pm $3.0 \\
\hline
$f_\text{OS} = 0.001$  &  0.090  &  0.465  & 117.6   \\
\scalebox{0.75}[0.75]{$\alpha_i=0.25 \rightarrow f_\text{OS} = 0.001$}$^a$ \hspace{-0.5cm} &  0.079 &  0.481  &  120.6 \\
\scalebox{0.75}[0.75]{$f_\text{OS} = 0.001 \rightarrow \alpha_i=0.25$}$^b$ \hspace{-0.5cm} &  0.102 &  0.429  &  112.2 \\
\hline
 Observations & 0.117$ \pm $0.005 & 0.455$ \pm $0.012 & \\
\hline
\multicolumn{4}{l}{$^a$ Sequence evolved with SOS ($\alpha_i=0.25$) until $Y_\text{cent} = 0.10$,} \\ \multicolumn{4}{l}{\hspace{0.2cm} and then evolved with standard overshoot.}\\
\multicolumn{4}{l}{$^b$ Sequence evolved with standard overshoot ($f_\text{OS} = 0.001$),} \\ \multicolumn{4}{l}{\hspace{0.2cm} until $Y_\text{cent} = 0.10$ and then evolved with SOS.}\\
\end{tabular}
\normalsize
\end{center}
\end{table}

\subsection{Core breathing pulses}
  \label{sec_cbp}

It is evident from our studies of globular cluster and solar-type stars in the previous sections that towards the end of CHeB the structures of the models with different overshooting prescriptions diverge.  In order to investigate whether the reduced occurrence of CBP with SOS is due to differences in the structure during the prior evolution, we calculated an evolution sequence where we switched to SOS near the end of a run that began with standard overshoot, but before the first CBP.  Changing the mixing scheme reduced the CHeB lifetime by 5\,Myr and increased $R_2$ from 0.090 to 0.102.  We also tested the effect of doing the opposite: beginning the evolution with SOS and finishing it with standard overshoot.  Switching to the standard overshoot scheme triggered a large breathing pulse that extended the CHeB lifetime by 11\,Myr and decreased $R_2$ from 0.103 to 0.079.  Both of these tests indicate that the dominant factor controlling the evolution is the mixing scheme late in CHeB, rather than the structure that develops in the earlier more quiescent phase.

\begin{figure}
\includegraphics[width=\linewidth]{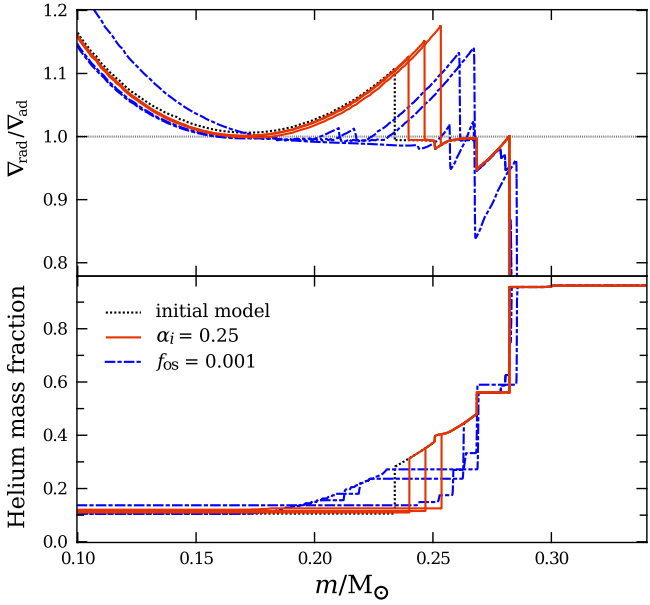}
  \caption{Evolution of the internal profile of $\nabla_\text{rad}/\nabla_\text{ad}$ during a core breathing pulse.  The initial 1\,$\text{M}_\odot$ model (black dotted line) was calculated with SOS and $\alpha_{i} = 0.25$ until the central helium mass fraction was $Y=0.106$ (and when the central helium abundance is increasing).  The evolution sequences are computed with SOS with $\alpha_{i} = 0.25$ (solid red lines) and standard overshoot with $f_\text{OS} = 0.001$ (blue dot-dash lines).  Lines are plotted at $5 \times 10^4$\,year intervals for the $\alpha_{i} = 0.25$ and at $10^4$\,year intervals for the standard overshoot model.}
  \label{figure_CBP_ratgrads}
\end{figure}

\begin{figure}
\includegraphics[width=\linewidth]{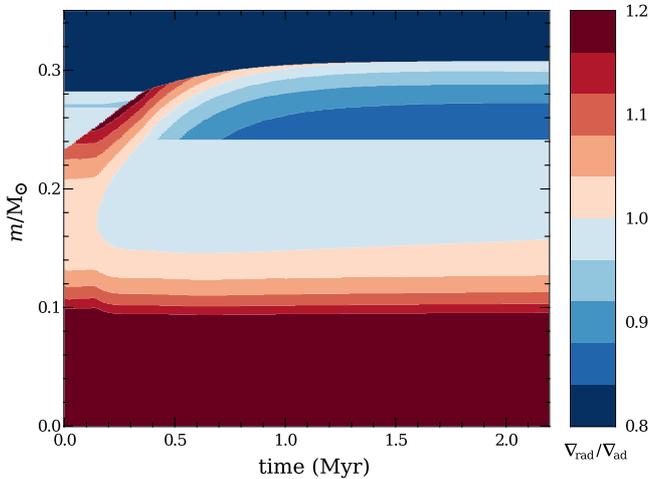}
  \caption{Kippenhahn plot showing the ratio of the temperature gradients $\nabla_\text{rad}/\nabla_\text{ad}$ during the core breathing pulse in the SOS sequence with $\alpha_{i} = 0.25$ from Figure~\ref{figure_CBP_ratgrads}.}
  \label{figure_CBP_kippenhahn}
\end{figure}

Now that we have established that the divergent evolution late in CHeB is driven by differences in the current mixing prescription rather than differences already imprinted onto the stellar structure, we analyse this in detail by computing two 1\,M$_\odot$ evolution sequences beginning from the same model: one with SOS and $\alpha_{i} = 0.25$ and another with standard overshoot and $f_\text{OS} = 0.001$.  The initial structure is taken from the evolution of a globular cluster model with SOS and $\alpha_{i} = 0.25$ until the central helium mass fraction is $Y_\text{cent}=0.106$ and it begins to increase. The subsequent evolution of $\nabla_\text{rad}/\nabla_\text{ad}$ and composition is shown in Figure~\ref{figure_CBP_ratgrads}.  Additionally, Figure~\ref{figure_CBP_kippenhahn} shows 2\,Myr of the evolution of $\nabla_\text{rad}/\nabla_\text{ad}$ for the SOS sequence.

In both the SOS and standard overshoot cases, mixing from overshoot initially increases the central helium abundance.  The contrasting rate of increase, however, is critical.  Although difficult to discern, the minimum in $\nabla_\text{rad}/\nabla_\text{ad}$ monotonically decreases slowly near $m = 0.16\,\text{M}_\odot$ (the location in the convection zone where it is lowest) for the model with $\alpha_{i} = 0.25$: it takes $10^5$\,yr for $\nabla_\text{rad}/\nabla_\text{ad}$ to decrease to unity, in which time the central helium abundance increases from $Y=0.106$ to $Y=0.121$.  In contrast, in the standard overshoot model, $\nabla_\text{rad}/\nabla_\text{ad}$ decreases to below unity and the convection zone splits nearly instantly.  The new convective shell then advances outward (in mass), leaving behind material with helium abundance increasing outward such that it is marginally stable, creating the composition profile that characterizes models with semiconvection \citep[e.g.][]{1972ApJ...171..309R}.  After $2 \times 10^4$\,yr in the standard overshoot sequence, when the core convection zone rapidly expands, the new reservoir of helium-rich material created by the advance of this convective shell is ingested, increasing the central helium abundance to $Y_\text{cent}=0.136$.

Regardless of the mixing scheme, the way that CBP terminate in these tests is no different from the long established picture \citep[e.g.][]{1971Ap&SS..10..355C,1985ApJ...296..204C}.  The value of the minimum value of $\nabla_\text{rad}/\nabla_\text{ad}$ inside the convection zone eventually decreases, which continues until $\nabla_\text{rad}/\nabla_\text{ad} < 1$, when the convection zone splits, halting the transport of helium into the convective core.  The rate of helium ingestion, however, is critical.  When it is fast, a strong feedback loop is initiated: the increased helium mass fraction temporarily boosts the helium-burning luminosity and increases $\nabla_\text{rad}$.  When the rate of ingestion is restricted, the luminosity increase is outweighed by the lower opacity and $\nabla_\text{rad}$ decreases, reducing the amount of helium ingested.  In order for the position of the convective boundary to be unstable, the rate of entrainment of helium into the core $\dot{m}_i$ must be faster than the rate it burns $\dot{m_b}$.  If we require $ \dot{m}_i> \dot{m_b} $ then Equation~16 in \citet{2015A&A...582L...2S} implies that 
\begin{equation}
\frac{\nabla_\text{rad}}{\nabla_\text{ad}} > \frac{\alpha_i \epsilon_b}{\alpha_i \epsilon_b-\frac{5}{12}RT}
\end{equation}
for the helium abundance in the convective core to increase, where $\epsilon_b$ is the specific energy released from helium burning.  In the model undergoing a breathing pulse in Figure~\ref{figure_CBP_ratgrads}, $\alpha_i \gtrsim 0.05$ is needed for the central helium abundance to increase.  This coincides with the disparate behaviour for the models with $\alpha_i < 0.05$ shown in Table~\ref{table_gc}, which have reduced $R_2$ and larger $\Delta \log{L}_\text{HB}^\text{AGB}$.

\subsection{Higher-mass models}

We also test the effect of SOS for models with higher initial stellar mass: 2\,M$_\odot$, 5\,M$_\odot$, and 10\,M$_\odot$.  The CHeB lifetimes for the suite of models, including some without any mixing beyond the Schwarzschild boundary (i.e. $\alpha_i = 0$) are shown in Table~\ref{table_initialmass}.  

Over the entire mass range, SOS significantly lengthens the CHeB lifetime compared with sequences without any convective overshoot (which is the case for any overshooting scheme).  In models with $M \leq 5\,\text{M}_\odot$, the inclusion of SOS shortens the CHeB lifetime compared with models with standard overshoot.  This effect, however, diminishes with increasing initial mass, and is not apparent for the 10$\,\text{M}_\odot$ model.  It is not surprising that the consequence of the mixing prescription around the convective core decreases for higher stellar mass because this corresponds to a growing contribution from hydrogen burning.

The efficiency parameter is least important to the 10\,M$_\odot$ sequences because none of the SOS models have CBP.  The standard overshoot sequence has only one small breathing pulse, which marginally increases the maximum mass enclosed by the convective core, to $m/\text{M}_\odot = 1.31$ compared with $m/\text{M}_\odot = 1.28$ for the $\alpha_i = 0.5$ model.  In that case, the secular expansion of the mass enclosed by the convective core shows the effect of the SOS efficiency: the higher $\alpha_i$, the closer the convection zone is to converging to a size such that $\nabla_\text{rad} = \nabla_\text{ad}$ at its boundary.  About three quarters of the way through CHeB, the boundary of the convective core is at $m/\text{M}_\odot = 1.000$, $1.042$, $1.068$, $1.076$, and $1.081$ for sequences with $\alpha_i = 0.05$, $0.10$, $0.25$, $0.5$, and $1.0$, respectively. This compares with $m/\text{M}_\odot = 1.100$ for the standard overshoot sequence.  

None of the 5\,M$_\odot$ sequences with SOS have CBP, and as a result their lifetime is less than the standard overshoot model, which has one breathing pulse.  Only two of the five 2\,M$_\odot$ models with SOS have a breathing pulse with $\Delta Y_\text{cent} > 0.05$.  In contrast, the standard overshoot model has three such CBP, increasing the CHeB lifetime by more than 11\,Myr compared with the longest lived SOS sequence.

\setlength{\tabcolsep}{4pt}
\begin{table}
\begin{center}
  \caption{Lifetime of CHeB phase (in Myr) for solar-metallicity models with different treatment of overshoot and initial mass.}
  \label{table_initialmass}
 \footnotesize
\begin{tabular}{cccccc}
\hline
$\alpha_i$ &  & 1\,$\text{M}_\odot$ & 2\,$\text{M}_\odot$ & 5\,$\text{M}_\odot$ & 10\,$\text{M}_\odot$ \\
\hline
$0.00$       &  &  83.31 &  61.30 & 11.21 & 2.34 \\
$0.05$       &  & 120.62 & 140.75 & 21.21 & 3.47 \\
$0.10$       &  & 116.26 & 138.88 & 21.51 & 3.58 \\
$0.25$       &  & 118.40 & 147.02 & 21.79 & 3.62 \\
$0.50$       &  & 121.17 & 139.82 & 21.32 & 3.62 \\
$1.00$       &  & 126.28 & 148.93 & 21.31 & 3.61 \\
\hline
$f_\text{os}=10^{-3}$ &  & 131.98 & 160.49 & 23.16 & 3.61 \\
\hline
\end{tabular}
\normalsize
\end{center}
\end{table}

\subsection{Applicability to other phases of evolution}

The constraint on the growth of convection zones that is explored in this paper is only relevant when there is a composition discontinuity and the convective luminosity does not vanish at the convective boundary.  Considering that $\nabla_\text{ad}$ only weakly depends on composition, this necessitates the radiative temperature gradient
\begin{equation}
\nabla_\text{rad} = \frac{3 \kappa P L }{64 \pi a c r^2 T^4 g}
\end{equation}
be discontinuous across the convective boundary.  Given that $P(r)$, $L(r)$, $T(r)$, and $g(r)$ must be continuous, the discontinuity must be in the opacity $\kappa$, and such that the opacity is higher inside the convection zone than outside.  This condition is satisfied in the CHeB case because carbon and oxygen are both more opaque than helium.  If, for example, this ingestion limit were applied to the well-studied case of main sequence convective core overshooting, no additional mixing beyond the Schwarzschild boundary would be permitted, whereas it is needed to conform with various lines of observational evidence, including observations of open clusters \citep{1991A&AS...89..451M}, eclipsing binaries \citep{2000MNRAS.318L..55R,2007A&A...475.1019C,2014ApJ...787..127M,2015A&A...575A.117S,2016A&A...592A..15C,2017A&A...600A..41V}, and asteroseismology \citep{2013ApJ...769..141S,2014ApJ...787..164G,2015MNRAS.453.2094Y,2016A&A...589A..93D,2016MNRAS.460.1254B}.  This suggests that there are other physical processes determining the location of the convective boundary in different physical regimes \citep[cf.][]{2015A&A...580A..61V}.

\section{Summary and conclusions}

In this study we examined the importance of the limit on the ingestion of helium into convective helium-burning cores recently proposed by \citet{2015A&A...582L...2S}.  This topic is of great interest because the evolution of core helium burning models is still subject to the uncertain mixing at the convective core boundary, and recently, novel asteroseismic observations have suggested there may be a fundamental problem with the models.  Using the Monash University stellar evolution code, we computed a suite of ``Spruit overshoot'' evolution sequences with a range of initial mass $0.83 \leq M/\text{M}_\odot \leq 10.0$ and overshooting efficiency $\alpha_i$.  We then assessed these models against the constraints from asteroseismology and globular cluster star counts that were outlined in the first two papers in this series.

The overall evolution of models with Spruit overshoot is relatively insensitive to the choice of $\alpha_i$ provided it is neither too low or too high, i.e. when $0.05 \lesssim \alpha_i \lesssim 1.0$.  For the most part, the evolution with Spruit overshoot is similar to that resulting from applying an exponential decay of the diffusion coefficient at each convective boundary (``standard overshoot'').  In the first half of CHeB, the growth of the convective core is consistent with sequences with standard overshoot.  Later in the evolution, however, the position of the convective core boundary does not become as unstable as it does in standard overshoot models.  Crucially, large episodes of helium ingestion, i.e. core breathing pulses, are suppressed, or in some cases, avoided entirely.  In all low mass ($M \leq 5\,\text{M}_\odot$) cases, reasonable values of the efficiency, i.e. $0.05 \leq \alpha_i \leq 1.0$, reduce the CHeB lifetime compared with standard overshoot models.

The reduction in the extent of core breathing pulses in the Spruit overshoot sequences minimizes several discrepancies that exist between observations of field and Galactic globular cluster stars and standard overshoot models.  Models with Spruit overshoot (and $\alpha_i \leq 1.0$) more closely match two constraints from star counts in globular clusters: $R_2$ and $\Delta\log{L}_\text{HB}^\text{AGB}$.  Additionally, because of the reduction of core breathing pulses, theoretical predictions for the luminosity probability density function resulting from models with Spruit overshoot have a sharper peak around the AGB clump, concordant with the observations.  Lastly, in the mode-trapping scenario posited in Paper~I to help reconcile models with asteroseismology, predictions from Spruit overshoot imply fewer stars with anomalously high g-mode period spacing near the end of core helium burning.

We conclude by stressing that we do not claim to present a complete theory: the constraint from \citet{2015A&A...582L...2S} involves a simplification that ignores non-local effects.  Instead, we have shown that substantial core breathing pulses are disfavoured, and a range of observations of core helium burning stars better matched, when accommodating this straightforward and physically motivated limitation on mixing.  In other phases, the situation appears more complex: an equivalent constraint is presumably violated during convective core hydrogen burning, when mixing beyond the Schwarzschild boundary is necessarily invoked to explain various observations.  We recommend further investigation into whether equivalent ingestion limits are applicable and important to other phases of stellar evolution and also whether the formulation used in this paper is consistent with results from state-of-the-art 3-dimensional hydrodynamical simulations \citep[e.g.][]{2016AN....337..788C}.

\section*{Acknowledgements}

The authors wish to acknowledge the contribution of Henk Spruit, who visited Monash University as part of the Monash Centre for Astrophysics Distinguished Visitor Program  in 2014.  This project was partly supported by the European Research Council through ERC AdG No. 320478-TOFU.

\footnotesize{
 \bibliographystyle{mnras}
  \bibliography{paper3_arxiv}

\begin{thebibliography}{}
\makeatletter
\relax
\def\mn@urlcharsother{\let\do\@makeother \do\$\do\&\do\#\do\^\do\_\do\%\do\~}
\def\mn@doi{\begingroup\mn@urlcharsother \@ifnextchar [ {\mn@doi@}
  {\mn@doi@[]}}
\def\mn@doi@[#1]#2{\def\@tempa{#1}\ifx\@tempa\@empty \href
  {http://dx.doi.org/#2} {doi:#2}\else \href {http://dx.doi.org/#2} {#1}\fi
  \endgroup}
\def\mn@eprint#1#2{\mn@eprint@#1:#2::\@nil}
\def\mn@eprint@arXiv#1{\href {http://arxiv.org/abs/#1} {{\tt arXiv:#1}}}
\def\mn@eprint@dblp#1{\href {http://dblp.uni-trier.de/rec/bibtex/#1.xml}
  {dblp:#1}}
\def\mn@eprint@#1:#2:#3:#4\@nil{\def\@tempa {#1}\def\@tempb {#2}\def\@tempc
  {#3}\ifx \@tempc \@empty \let \@tempc \@tempb \let \@tempb \@tempa \fi \ifx
  \@tempb \@empty \def\@tempb {arXiv}\fi \@ifundefined
  {mn@eprint@\@tempb}{\@tempb:\@tempc}{\expandafter \expandafter \csname
  mn@eprint@\@tempb\endcsname \expandafter{\@tempc}}}

\bibitem[\protect\citeauthoryear{{Bazot}, {Christensen-Dalsgaard}, {Gizon}  \&
  {Benomar}}{{Bazot} et~al.}{2016}]{2016MNRAS.460.1254B}
{Bazot} M.,  {Christensen-Dalsgaard} J.,  {Gizon} L.,   {Benomar} O.,  2016,
  \mn@doi [\mnras] {10.1093/mnras/stw921}, \href
  {http://adsabs.harvard.edu/abs/2016MNRAS.460.1254B} {460, 1254}

\bibitem[\protect\citeauthoryear{{Bedding} et~al.,}{{Bedding}
  et~al.}{2010}]{2010ApJ...713L.176B}
{Bedding} T.~R.,  et~al., 2010, \mn@doi [\apjl] {10.1088/2041-8205/713/2/L176},
  \href {http://adsabs.harvard.edu/abs/2010ApJ...713L.176B} {713, L176}

\bibitem[\protect\citeauthoryear{{Bono}, {Caputo}, {Cassisi}, {Castellani}  \&
  {Marconi}}{{Bono} et~al.}{1997a}]{1997ApJ...479..279B}
{Bono} G.,  {Caputo} F.,  {Cassisi} S.,  {Castellani} V.,   {Marconi} M.,
  1997a, \apj, \href {http://adsabs.harvard.edu/abs/1997ApJ...479..279B} {479,
  279}

\bibitem[\protect\citeauthoryear{{Bono}, {Caputo}, {Cassisi}, {Castellani}  \&
  {Marconi}}{{Bono} et~al.}{1997b}]{1997ApJ...489..822B}
{Bono} G.,  {Caputo} F.,  {Cassisi} S.,  {Castellani} V.,   {Marconi} M.,
  1997b, \apj, \href {http://adsabs.harvard.edu/abs/1997ApJ...489..822B} {489,
  822}

\bibitem[\protect\citeauthoryear{Bossini}{Bossini}{2016}]{etheses7090}
Bossini D.,  2016, PhD thesis, University of Birmingham

\bibitem[\protect\citeauthoryear{{Bossini} et~al.,}{{Bossini}
  et~al.}{2015}]{2015MNRAS.453.2290B}
{Bossini} D.,  et~al., 2015, \mn@doi [\mnras] {10.1093/mnras/stv1738}, \href
  {http://ukads.nottingham.ac.uk/abs/2015MNRAS.453.2290B} {453, 2290}

\bibitem[\protect\citeauthoryear{{Bossini} et~al.,}{{Bossini}
  et~al.}{2017}]{2017arXiv170503077B}
{Bossini} D.,  et~al., 2017, preprint, \href
  {http://adsabs.harvard.edu/abs/2017arXiv170503077B} {} (\mn@eprint {arXiv}
  {1705.03077})

\bibitem[\protect\citeauthoryear{{Bressan}, {Bertelli}  \& {Chiosi}}{{Bressan}
  et~al.}{1986}]{1986MmSAI..57..411B}
{Bressan} A.,  {Bertelli} G.,   {Chiosi} C.,  1986, \memsai, \href
  {http://adsabs.harvard.edu/abs/1986MmSAI..57..411B} {57, 411}

\bibitem[\protect\citeauthoryear{{Caloi} \& {Mazzitelli}}{{Caloi} \&
  {Mazzitelli}}{1990}]{1990A&A...240..305C}
{Caloi} V.,  {Mazzitelli} I.,  1990, \aap, \href
  {http://adsabs.harvard.edu/abs/1990A%26A...240..305C} {240, 305}

\bibitem[\protect\citeauthoryear{{Campbell} \& {Lattanzio}}{{Campbell} \&
  {Lattanzio}}{2008}]{2008A&A...490..769C}
{Campbell} S.~W.,  {Lattanzio} J.~C.,  2008, \mn@doi [\aap]
  {10.1051/0004-6361:200809597}, \href
  {http://adsabs.harvard.edu/abs/2008A%26A...490..769C} {490, 769}

\bibitem[\protect\citeauthoryear{{Campbell} et~al.,}{{Campbell}
  et~al.}{2016}]{2016AN....337..788C}
{Campbell} S.~W.,  et~al., 2016, \mn@doi [Astronomische Nachrichten]
  {10.1002/asna.201612373}, \href
  {http://adsabs.harvard.edu/abs/2016AN....337..788C} {337, 788}

\bibitem[\protect\citeauthoryear{{Caputo}, {Chieffi}, {Tornambe}, {Castellani}
  \& {Pulone}}{{Caputo} et~al.}{1989}]{1989ApJ...340..241C}
{Caputo} F.,  {Chieffi} A.,  {Tornambe} A.,  {Castellani} V.,   {Pulone} L.,
  1989, \mn@doi [\apj] {10.1086/167387}, \href
  {http://adsabs.harvard.edu/abs/1989ApJ...340..241C} {340, 241}

\bibitem[\protect\citeauthoryear{{Cassisi}, {Castellani}, {Degl'Innocenti},
  {Piotto}  \& {Salaris}}{{Cassisi} et~al.}{2001}]{2001A&A...366..578C}
{Cassisi} S.,  {Castellani} V.,  {Degl'Innocenti} S.,  {Piotto} G.,   {Salaris}
  M.,  2001, \mn@doi [\aap] {10.1051/0004-6361:20000293}, \href
  {http://adsabs.harvard.edu/abs/2001A%26A...366..578C} {366, 578}

\bibitem[\protect\citeauthoryear{{Castellani}, {Giannone}  \&
  {Renzini}}{{Castellani} et~al.}{1971a}]{1971Ap&SS..10..340C}
{Castellani} V.,  {Giannone} P.,   {Renzini} A.,  1971a, \mn@doi [\apss]
  {10.1007/BF00704092}, \href
  {http://adsabs.harvard.edu/abs/1971Ap%26SS..10..340C} {10, 340}

\bibitem[\protect\citeauthoryear{{Castellani}, {Giannone}  \&
  {Renzini}}{{Castellani} et~al.}{1971b}]{1971Ap&SS..10..355C}
{Castellani} V.,  {Giannone} P.,   {Renzini} A.,  1971b, \mn@doi [\apss]
  {10.1007/BF00649680}, \href
  {http://adsabs.harvard.edu/abs/1971Ap%26SS..10..355C} {10, 355}

\bibitem[\protect\citeauthoryear{{Castellani}, {Chieffi}, {Tornambe}  \&
  {Pulone}}{{Castellani} et~al.}{1985}]{1985ApJ...296..204C}
{Castellani} V.,  {Chieffi} A.,  {Tornambe} A.,   {Pulone} L.,  1985, \mn@doi
  [\apj] {10.1086/163437}, \href
  {http://adsabs.harvard.edu/abs/1985ApJ...296..204C} {296, 204}

\bibitem[\protect\citeauthoryear{{Claret}}{{Claret}}{2007}]{2007A&A...475.1019C}
{Claret} A.,  2007, \mn@doi [\aap] {10.1051/0004-6361:20078024}, \href
  {http://adsabs.harvard.edu/abs/2007A%26A...475.1019C} {475, 1019}

\bibitem[\protect\citeauthoryear{{Claret} \& {Torres}}{{Claret} \&
  {Torres}}{2016}]{2016A&A...592A..15C}
{Claret} A.,  {Torres} G.,  2016, \mn@doi [\aap] {10.1051/0004-6361/201628779},
  \href {http://adsabs.harvard.edu/abs/2016A%26A...592A..15C} {592, A15}

\bibitem[\protect\citeauthoryear{{Constantino}, {Campbell}, {Gil-Pons}  \&
  {Lattanzio}}{{Constantino} et~al.}{2014}]{2014ApJ...784...56C}
{Constantino} T.,  {Campbell} S.,  {Gil-Pons} P.,   {Lattanzio} J.,  2014,
  \mn@doi [\apj] {10.1088/0004-637X/784/1/56}, \href
  {http://cdsads.u-strasbg.fr/abs/2014ApJ...784...56C} {784, 56}

\bibitem[\protect\citeauthoryear{{Constantino}, {Campbell},
  {Christensen-Dalsgaard}, {Lattanzio}  \& {Stello}}{{Constantino}
  et~al.}{2015}]{2015MNRAS.452..123C}
{Constantino} T.,  {Campbell} S.~W.,  {Christensen-Dalsgaard} J.,  {Lattanzio}
  J.~C.,   {Stello} D.,  2015, \mn@doi [\mnras] {10.1093/mnras/stv1264}, \href
  {http://adsabs.harvard.edu/abs/2015MNRAS.452..123C} {452, 123}

\bibitem[\protect\citeauthoryear{{Constantino}, {Campbell}, {Lattanzio}  \&
  {van Duijneveldt}}{{Constantino} et~al.}{2016}]{2016MNRAS.456.3866C}
{Constantino} T.,  {Campbell} S.~W.,  {Lattanzio} J.~C.,   {van Duijneveldt}
  A.,  2016, \mn@doi [\mnras] {10.1093/mnras/stv2939}, \href
  {http://adsabs.harvard.edu/abs/2016MNRAS.456.3866C} {456, 3866}

\bibitem[\protect\citeauthoryear{{Deheuvels}, {Brand{\~a}o}, {Silva Aguirre},
  {Ballot}, {Michel}, {Cunha}, {Lebreton}  \& {Appourchaux}}{{Deheuvels}
  et~al.}{2016}]{2016A&A...589A..93D}
{Deheuvels} S.,  {Brand{\~a}o} I.,  {Silva Aguirre} V.,  {Ballot} J.,  {Michel}
  E.,  {Cunha} M.~S.,  {Lebreton} Y.,   {Appourchaux} T.,  2016, \mn@doi [\aap]
  {10.1051/0004-6361/201527967}, \href
  {http://adsabs.harvard.edu/abs/2016A%26A...589A..93D} {589, A93}

\bibitem[\protect\citeauthoryear{{Demarque} \& {Mengel}}{{Demarque} \&
  {Mengel}}{1972}]{1972ApJ...171..583D}
{Demarque} P.,  {Mengel} J.~G.,  1972, \mn@doi [\apj] {10.1086/151312}, \href
  {http://adsabs.harvard.edu/abs/1972ApJ...171..583D} {171, 583}

\bibitem[\protect\citeauthoryear{{Denissenkov}, {VandenBerg}, {Kopacki}  \&
  {Ferguson}}{{Denissenkov} et~al.}{2017}]{2017arXiv170605454D}
{Denissenkov} P.,  {VandenBerg} D.~A.,  {Kopacki} G.,   {Ferguson} J.~W.,
  2017, preprint, \href {http://adsabs.harvard.edu/abs/2017arXiv170605454D} {}
  (\mn@eprint {arXiv} {1706.05454})

\bibitem[\protect\citeauthoryear{{Dorman} \& {Rood}}{{Dorman} \&
  {Rood}}{1993}]{1993ApJ...409..387D}
{Dorman} B.,  {Rood} R.~T.,  1993, \mn@doi [\apj] {10.1086/172671}, \href
  {http://adsabs.harvard.edu/abs/1993ApJ...409..387D} {409, 387}

\bibitem[\protect\citeauthoryear{{Dziembowski}}{{Dziembowski}}{1977}]{1977AcA....27...95D}
{Dziembowski} W.,  1977, \actaa, \href
  {http://adsabs.harvard.edu/abs/1977AcA....27...95D} {27, 95}

\bibitem[\protect\citeauthoryear{{Gabriel}, {Noels}, {Montalb{\'a}n}  \&
  {Miglio}}{{Gabriel} et~al.}{2014}]{2014A&A...569A..63G}
{Gabriel} M.,  {Noels} A.,  {Montalb{\'a}n} J.,   {Miglio} A.,  2014, \mn@doi
  [\aap] {10.1051/0004-6361/201423442}, \href
  {http://adsabs.harvard.edu/abs/2014A%26A...569A..63G} {569, A63}

\bibitem[\protect\citeauthoryear{{Ghasemi}, {Moravveji}, {Aerts}, {Safari}  \&
  {Vu{\v c}kovi{\'c}}}{{Ghasemi} et~al.}{2017}]{2017MNRAS.465.1518G}
{Ghasemi} H.,  {Moravveji} E.,  {Aerts} C.,  {Safari} H.,   {Vu{\v c}kovi{\'c}}
  M.,  2017, \mn@doi [\mnras] {10.1093/mnras/stw2839}, \href
  {http://adsabs.harvard.edu/abs/2017MNRAS.465.1518G} {465, 1518}

\bibitem[\protect\citeauthoryear{{Girardi}}{{Girardi}}{1999}]{1999MNRAS.308..818G}
{Girardi} L.,  1999, \mn@doi [\mnras] {10.1046/j.1365-8711.1999.02746.x}, \href
  {http://adsabs.harvard.edu/abs/1999MNRAS.308..818G} {308, 818}

\bibitem[\protect\citeauthoryear{{Guenther}, {Demarque}  \&
  {Gruberbauer}}{{Guenther} et~al.}{2014}]{2014ApJ...787..164G}
{Guenther} D.~B.,  {Demarque} P.,   {Gruberbauer} M.,  2014, \mn@doi [\apj]
  {10.1088/0004-637X/787/2/164}, \href
  {http://adsabs.harvard.edu/abs/2014ApJ...787..164G} {787, 164}

\bibitem[\protect\citeauthoryear{{Herwig}, {Bloecker}, {Schoenberner}  \& {El
  Eid}}{{Herwig} et~al.}{1997}]{1997A&A...324L..81H}
{Herwig} F.,  {Bloecker} T.,  {Schoenberner} D.,   {El Eid} M.,  1997, \aap,
  \href {http://adsabs.harvard.edu/abs/1997A%26A...324L..81H} {324, L81}

\bibitem[\protect\citeauthoryear{{Lattanzio}}{{Lattanzio}}{1986}]{1986ApJ...311..708L}
{Lattanzio} J.~C.,  1986, \mn@doi [\apj] {10.1086/164810}, \href
  {http://adsabs.harvard.edu/abs/1986ApJ...311..708L} {311, 708}

\bibitem[\protect\citeauthoryear{{Maeder} \& {Meynet}}{{Maeder} \&
  {Meynet}}{1991}]{1991A&AS...89..451M}
{Maeder} A.,  {Meynet} G.,  1991, \aaps, \href
  {http://adsabs.harvard.edu/abs/1991A%26AS...89..451M} {89, 451}

\bibitem[\protect\citeauthoryear{{Meng} \& {Zhang}}{{Meng} \&
  {Zhang}}{2014}]{2014ApJ...787..127M}
{Meng} Y.,  {Zhang} Q.~S.,  2014, \mn@doi [\apj] {10.1088/0004-637X/787/2/127},
  \href {http://adsabs.harvard.edu/abs/2014ApJ...787..127M} {787, 127}

\bibitem[\protect\citeauthoryear{{Miglio}, {Montalb{\'a}n}, {Noels}  \&
  {Eggenberger}}{{Miglio} et~al.}{2008}]{2008MNRAS.386.1487M}
{Miglio} A.,  {Montalb{\'a}n} J.,  {Noels} A.,   {Eggenberger} P.,  2008,
  \mn@doi [\mnras] {10.1111/j.1365-2966.2008.13112.x}, \href
  {http://esoads.eso.org/abs/2008MNRAS.386.1487M} {386, 1487}

\bibitem[\protect\citeauthoryear{{Montalb{\'a}n}, {Miglio}, {Noels}, {Dupret},
  {Scuflaire}  \& {Ventura}}{{Montalb{\'a}n}
  et~al.}{2013}]{2013ApJ...766..118M}
{Montalb{\'a}n} J.,  {Miglio} A.,  {Noels} A.,  {Dupret} M.-A.,  {Scuflaire}
  R.,   {Ventura} P.,  2013, \mn@doi [\apj] {10.1088/0004-637X/766/2/118},
  \href {http://adsabs.harvard.edu/abs/2013ApJ...766..118M} {766, 118}

\bibitem[\protect\citeauthoryear{{Mosser} et~al.,}{{Mosser}
  et~al.}{2012}]{2012A&A...540A.143M}
{Mosser} B.,  et~al., 2012, \mn@doi [\aap] {10.1051/0004-6361/201118519}, \href
  {http://adsabs.harvard.edu/abs/2012A%26A...540A.143M} {540, A143}

\bibitem[\protect\citeauthoryear{{Mosser} et~al.,}{{Mosser}
  et~al.}{2014}]{2014A&A...572L...5M}
{Mosser} B.,  et~al., 2014, \mn@doi [\aap] {10.1051/0004-6361/201425039}, \href
  {http://adsabs.harvard.edu/abs/2014A%26A...572L...5M} {572, L5}

\bibitem[\protect\citeauthoryear{{Mosser} et~al.,}{{Mosser}
  et~al.}{2017}]{2017A&A...598A..62M}
{Mosser} B.,  et~al., 2017, \mn@doi [\aap] {10.1051/0004-6361/201629494}, \href
  {http://adsabs.harvard.edu/abs/2017A%26A...598A..62M} {598, A62}

\bibitem[\protect\citeauthoryear{{Pinsonneault} et~al.,}{{Pinsonneault}
  et~al.}{2014}]{2014ApJS..215...19P}
{Pinsonneault} M.~H.,  et~al., 2014, \mn@doi [\apjs]
  {10.1088/0067-0049/215/2/19}, \href
  {http://adsabs.harvard.edu/abs/2014ApJS..215...19P} {215, 19}

\bibitem[\protect\citeauthoryear{{Reimers}}{{Reimers}}{1975}]{1975MSRSL...8..369R}
{Reimers} D.,  1975, Memoires of the Societe Royale des Sciences de Liege,
  \href {http://adsabs.harvard.edu/abs/1975MSRSL...8..369R} {8, 369}

\bibitem[\protect\citeauthoryear{{Ribas}, {Jordi}  \& {Gim{\'e}nez}}{{Ribas}
  et~al.}{2000}]{2000MNRAS.318L..55R}
{Ribas} I.,  {Jordi} C.,   {Gim{\'e}nez} {\'A}.,  2000, \mn@doi [\mnras]
  {10.1046/j.1365-8711.2000.04035.x}, \href
  {http://adsabs.harvard.edu/abs/2000MNRAS.318L..55R} {318, L55}

\bibitem[\protect\citeauthoryear{{Robertson} \& {Faulkner}}{{Robertson} \&
  {Faulkner}}{1972}]{1972ApJ...171..309R}
{Robertson} J.~W.,  {Faulkner} D.~J.,  1972, \mn@doi [\apj] {10.1086/151283},
  \href {http://adsabs.harvard.edu/abs/1972ApJ...171..309R} {171, 309}

\bibitem[\protect\citeauthoryear{{Schindler}, {Green}  \& {Arnett}}{{Schindler}
  et~al.}{2015}]{2015ApJ...806..178S}
{Schindler} J.-T.,  {Green} E.~M.,   {Arnett} W.~D.,  2015, \mn@doi [\apj]
  {10.1088/0004-637X/806/2/178}, \href
  {http://adsabs.harvard.edu/abs/2015ApJ...806..178S} {806, 178}

\bibitem[\protect\citeauthoryear{{Silva Aguirre} et~al.,}{{Silva Aguirre}
  et~al.}{2013}]{2013ApJ...769..141S}
{Silva Aguirre} V.,  et~al., 2013, \mn@doi [\apj]
  {10.1088/0004-637X/769/2/141}, \href
  {http://adsabs.harvard.edu/abs/2013ApJ...769..141S} {769, 141}

\bibitem[\protect\citeauthoryear{{Spruit}}{{Spruit}}{2015}]{2015A&A...582L...2S}
{Spruit} H.~C.,  2015, \mn@doi [\aap] {10.1051/0004-6361/201527171}, \href
  {http://adsabs.harvard.edu/abs/2015A%26A...582L...2S} {582, L2}

\bibitem[\protect\citeauthoryear{{Stancliffe}, {Fossati}, {Passy}  \&
  {Schneider}}{{Stancliffe} et~al.}{2015}]{2015A&A...575A.117S}
{Stancliffe} R.~J.,  {Fossati} L.,  {Passy} J.-C.,   {Schneider} F.~R.~N.,
  2015, \mn@doi [\aap] {10.1051/0004-6361/201425126}, \href
  {http://adsabs.harvard.edu/abs/2015A%26A...575A.117S} {575, A117}

\bibitem[\protect\citeauthoryear{{Sweigart}}{{Sweigart}}{1991}]{1991ASPC...13..299S}
{Sweigart} A.~V.,  1991, in {Janes} K.,  ed.,  Astronomical Society of the
  Pacific Conference Series Vol. 13, The Formation and Evolution of Star
  Clusters. pp 299--301

\bibitem[\protect\citeauthoryear{{Sweigart} \& {Demarque}}{{Sweigart} \&
  {Demarque}}{1973}]{1973ASSL...36..221S}
{Sweigart} A.~V.,  {Demarque} P.,  1973, in {Fernie} J.~D.,  ed.,  Astrophysics
  and Space Science Library Vol. 36, IAU Colloq. 21: Variable Stars in Globular
  Clusters and in Related Systems. p.~221

\bibitem[\protect\citeauthoryear{{Sweigart}, {Lattanzio}, {Gray}  \&
  {Tout}}{{Sweigart} et~al.}{2000}]{2000LIACo..35..529S}
{Sweigart} A.~V.,  {Lattanzio} J.~C.,  {Gray} J.~P.,   {Tout} C.~A.,  2000, in
  {Noels} A.,  {Magain} P.,  {Caro} D.,  {Jehin} E.,  {Parmentier} G.,
  {Thoul} A.~A.,  eds,  Liege International Astrophysical Colloquia Vol. 35,
  Liege International Astrophysical Colloquia. p.~529 (\mn@eprint {}
  {astro-ph/9909404})

\bibitem[\protect\citeauthoryear{{Valle}, {Dell'Omodarme}, {Prada Moroni}  \&
  {Degl'Innocenti}}{{Valle} et~al.}{2017}]{2017A&A...600A..41V}
{Valle} G.,  {Dell'Omodarme} M.,  {Prada Moroni} P.~G.,   {Degl'Innocenti} S.,
  2017, \mn@doi [\aap] {10.1051/0004-6361/201628240}, \href
  {http://adsabs.harvard.edu/abs/2017A%26A...600A..41V} {600, A41}

\bibitem[\protect\citeauthoryear{{VandenBerg}, {Denissenkov}  \&
  {Catelan}}{{VandenBerg} et~al.}{2016}]{2016ApJ...827....2V}
{VandenBerg} D.~A.,  {Denissenkov} P.~A.,   {Catelan} M.,  2016, \mn@doi [\apj]
  {10.3847/0004-637X/827/1/2}, \href
  {http://adsabs.harvard.edu/abs/2016ApJ...827....2V} {827, 2}

\bibitem[\protect\citeauthoryear{{Viallet}, {Meakin}, {Prat}  \&
  {Arnett}}{{Viallet} et~al.}{2015}]{2015A&A...580A..61V}
{Viallet} M.,  {Meakin} C.,  {Prat} V.,   {Arnett} D.,  2015, \mn@doi [\aap]
  {10.1051/0004-6361/201526294}, \href
  {http://adsabs.harvard.edu/abs/2015A%26A...580A..61V} {580, A61}

\bibitem[\protect\citeauthoryear{{Vrard}, {Mosser}  \& {Samadi}}{{Vrard}
  et~al.}{2016}]{2016A&A...588A..87V}
{Vrard} M.,  {Mosser} B.,   {Samadi} R.,  2016, \mn@doi [\aap]
  {10.1051/0004-6361/201527259}, \href
  {http://adsabs.harvard.edu/abs/2016A%26A...588A..87V} {588, A87}

\bibitem[\protect\citeauthoryear{{Yang}, {Tian}, {Bi}, {Ge}, {Wu}  \&
  {Zhang}}{{Yang} et~al.}{2015}]{2015MNRAS.453.2094Y}
{Yang} W.,  {Tian} Z.,  {Bi} S.,  {Ge} Z.,  {Wu} Y.,   {Zhang} J.,  2015,
  \mn@doi [\mnras] {10.1093/mnras/stv1841}, \href
  {http://adsabs.harvard.edu/abs/2015MNRAS.453.2094Y} {453, 2094}

\makeatother
\end{thebibliography}
}

\end{document}